\documentclass[preprint]{aastex}

\shorttitle{Fine Structure Lines of Hydrogen}
\shortauthors{Dennison et al.}

\begin{document}

\title{The Fine Structure Lines of Hydrogen in HII Regions}

\author{Brian Dennison}
\affil{Physics Department, 
and Pisgah Astronomical Research and Science Education Center, 
University of North Carolina at Asheville, Asheville, NC 28804}
\email{dennison@unca.edu}

\author{B. E. Turner}
\affil{National Radio Astronomy Observatory, 520 Edgement Road, Charlottesville, VA 22903-2475}
\email{bturner@nrao.edu}

\author{Anthony H. Minter}
\affil{National Radio Astronomy Observatory, P. O. Box 2, Green Bank, WV 24944}
\email{tminter@nrao.edu}

\begin{abstract}

The $2s_{1/2}$ state of hydrogen is metastable and overpopulated in HII regions.  In addition, the $2p$ states may be pumped by ambient Lyman-$\alpha$ radiation.  Fine structure transitions between these states may be observable in HII regions at 1.1 GHz ($2s_{1/2}-2p_{1/2}$) and/or 9.9 GHz ($2s_{1/2}-2p_{3/2}$), although the details of absorption versus emission are determined by the relative populations of the $2s$ and $2p$ states.  The $n=2$ level populations are solved with a parameterization that allows for Lyman-$\alpha$ pumping of the $2p$ states.  The Lyman-$\alpha$ pumping rate has long been considered uncertain as it involves solution of the difficult Lyman-$\alpha$ transfer problem.  The density of Lyman-$\alpha$ photons is set by their creation rate, easily determined from the recombination rate, and their removal rate.  Here we suggest that the dominant removal mechanism of Lyman-$\alpha$ radiation in HII regions is absorption by dust.  This circumvents the need to solve the Lyman-$\alpha$ transfer problem, and provides an upper limit to the rate at which the $2p$ states are populated by Lyman-$\alpha$ photons.  In virtually all cases of interest, the $2p$ states are predominantly populated by recombination, rather than Lyman-$\alpha$ pumping.  We then solve the radiative transfer problem for the fine structure lines in the presence of free-free radiation.  In the likely absence of Lyman-$\alpha$ pumping, the $2s_{1/2}\rightarrow 2p_{1/2}$ lines will appear in stimulated emission and the $2s_{1/2}\rightarrow 2p_{3/2}$ lines in absorption.  Because the final $2p$ states are short lived these lines are dominated by intrinsic linewidth (99.8 MHz).  In addition, each fine structure line is a multiplet of three blended hyperfine transitions.  Searching for the 9.9 GHz lines in high emission measure HII regions offers the best prospects for detection.  The lines are predicted to be weak; in the best cases, line-to continuum ratios of several tenths of a percent might be expected with line strengths of tens to a hundred mK with the Green Bank Telescope.  Predicted line strengths, at both 1.1 and 9.9 GHz, are given for a number of HII regions, high emission measure components, and planetary nebulae, based upon somewhat uncertain emission measures, sizes, and structures.  The extraordinary width of these lines and their blended structure will complicate detection.

\end{abstract}



\section{Introduction}

Transitions between fine-structure sublevels in atomic hydrogen have
never been detected astronomically. With advances in radio astronomy,
however, it now appears that transitions between the $2s_{1/2}$ and
$2p_{1/2}$, and $2s_{1/2}$ and $2 p_{3/2}$ levels may soon be observable
in HII regions. Figure 1 shows the fine and hyperfine structure of the
$n=2$ level of atomic hydrogen. The fine structure splitting into the
$2s_{1/2}$, $2p_{1/2}$, and $2p_{3/2}$ levels is caused by a combination
of spin orbit coupling, relativistic effects and the Lamb shift.
Significantly, the $2s_{1/2}$ state is metastable owing to the transition
rules for angular momentum, i.e. the $2s$--$1s$ transition is forbidden.
Hydrogen atoms in the interstellar medium are removed from
the $2s_{1/2}$ state principally through 2-photon emission to the ground
state (Breit and Teller 1940) with decay rate per atom, $A_{2\gamma} =
8.227\ {\rm sec}^{-1}$ (Spitzer and Greenstein 1951), resulting in
optical continuum radiation. Because this rate is some eight orders of
magnitude slower than Lyman decay, it may seem reasonable to expect that
the $2s_{1/2}$ state will be overpopulated relative to other excited
states under photoionization equilibrium. It must be noted, however,
that trapped Lyman-$\alpha$ radiation in an HII region will pump the
$2p$ states. In addition, in dense HII regions collisions, primarily
with ions, will transfer hydrogen atoms between the $2s$ and $2p$
states. These effects must be taken into account in order to determine
the expected strengths and signs (absorption versus stimulated emission)
of the $2s_{1/2}$-$2p_{1/2}$ (1.1 GHz) and $2s_{1/2}$-$2p_{3/2}$ (9.9
GHz) lines.

\begin{figure}
\epsscale{0.7}
\plotone{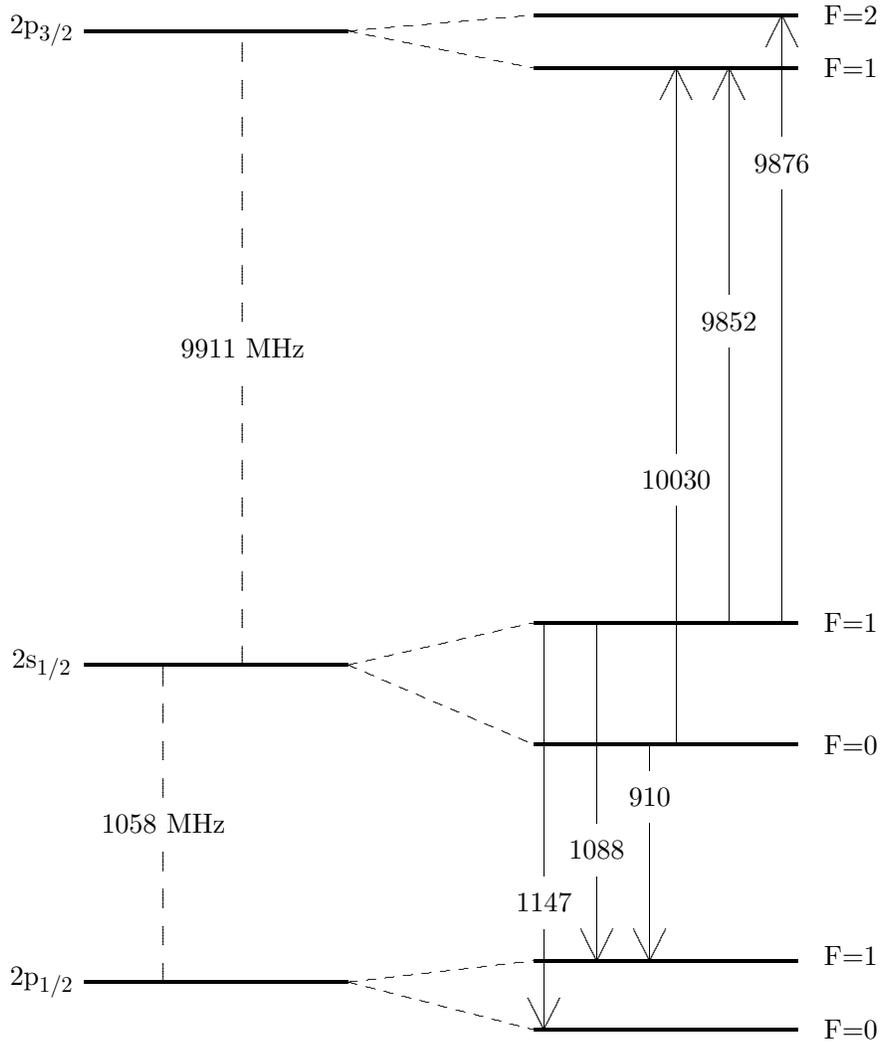}
\caption{Fine and hyperfine structure of the $n=2$ level of hydrogen.  Allowed hyperfine transitions are shown as solid vertical lines, with arrows denoting predominant transition directions if the metastable $2s_{1/2}$ states are overpopulated with respect to the $2p$ states.  (See text.)  Hyperfine transition frequencies are given in MHz.  The quantum number $F$ corresponds to total angular momentum, i.e., orbit + electron spin + nuclear spin.  Note that level separations are not shown to scale.}
\end{figure}

Apparently, \citet{wil52} was one of the first to investigate the
astrophysical implications of these lines, suggesting that the
$2s_{1/2}$-$2p_{3/2}$ line might be observed in the sun. \citet{pur52},
however, noted that collisions in the dense solar atmosphere would
equilibrate the $2s$ and $2p$ populations making detection unlikely.
\citet{tow57} suggested that an overpopulation of the metastable $2s$
states may lead to detectable lines in the interstellar medium.
\citet{shk60} raised the possibility of Lyman-$\alpha$ pumping of the
$2p$ states but discounted its sufficiency for producing observable
$2p\rightarrow 2s$ transitions. \citet{pot60} argued that the
Lyman-$\alpha$ depth in HII regions is essentially infinite such that
virtually all downward Lyman-$\alpha$ transitions are balanced by a
Lyman-$\alpha$ absorption, resulting in an overpopulation of $2p$ states
relative to $2s$ at densities below about $10^4$ cm$^{-3}$. \citet{fie61} 
further considered the case in which Lyman-$\alpha$
radiation is effectively destroyed by collisional conversion of $2p$
states to $2s$ (followed by 2-photon decay to the ground state). The
consequent overpopulation of $2p$ states then led to the prediction that
the 9.9 GHz lines would appear in stimulated emission in HII regions.
(This also implied that the 1.1 GHz lines would appear in absorption.)
\citet{mye72} used the Haystack 37-m radiotelescope with a
16-channel filter bank spanning 1 GHz in frequency to search for the 9.9
GHz lines, but did not find any lines in excess of 0.1 K antenna
temperature.  \citet{ers87} estimated the strengths of both sets of 
lines in Orion A and W3(OH) under the assumption that the $2p$ states 
are negligibly populated with respect to $2s_{1/2}$.

The major variation in the predictions arises from the
uncertain density of Lyman-$\alpha$ radiation in HII regions, which in
turn is determined by the processes that destroy Lyman-$\alpha$
radiation.  In addition to the conversion of $2p$ to $2s$ states via
collisions, other processes are likely to be important.  Resonantly
scattered Lyman-$\alpha$ photons undergo a diffusion in frequency 
and may thus acquire a significant probability of escaping the HII
region due the much lower optical depth in the line wings \citep{cox69,spi78}.  
Systematic gas motions within the region
(e.g. expansion) can also contribute to the migration of Lyman-$\alpha$
photons away from line center.  The dominant removal mechanism, however,
is absorption by dust within the HII region \citep{kap70,spi78}.  This last
fact greatly simplifies the estimation of the Lyman-$\alpha$ density.

In this paper, we obtain formulae for the $2s$ and $2p$ populations in an HII region under a parameterization that characterizes the rate of $2p$ production by Lyman-$\alpha$ absorption in terms of the rate of $2p$ production through recombination (Section 2).  Existing models for dust in HII regions and planetary nebulae are then used to place tight constraints on the Lyman-$\alpha$ density and thus the $2p$ production rate by Lyman-$\alpha$ photons (Section 3).  The radiative transfer problem for the fine structure lines is then solved for a uniform region (Section 4).  We obtain approximate predicted line strengths for various HII regions, compact components, and planetary nebulae under the justified assumption that Lyman-$\alpha$ pumping of the $2p$ states is negligible (Section 5). 
High emission measure HII regions and/or planetary 
nebulae are most likely to
yield detectable lines.

\section{Population of the $2s$ and $2p$ States}

\citet{spi51} estimated that between 0.30 and 0.35 of
all recombinations in an HII region reach the $2s$ state.
Including the effects of collisions and the 2-photon decay rate, the rate equation for the $2s$ state is
\begin{equation}
f\alpha n_i n_e + C_{ps}n_i n_{2p} = A_{2\gamma}n_{2s} + C_{sp}n_i n_{2s} ,
\end{equation}
where $f$ is the fraction of recombinations producing the $2s$
state and $\alpha$ is the recombination coefficient.  The collision
coefficients, $C_{ps}$ and $C_{sp}$, give the appropriate rates at which atoms are transferred from $2p$ to $2s$
and from $2s$ to $2p$, respectively.  These rates are
dominated by collisions with protons, but including electron collisions
(and assuming that $n_e = n_i$ and $T = 10^4$ K), we have $C_{sp} = 5.31
\times 10^{-4}$ cm$^3$ s$^{-1}$ \citep{sea55}.  Since $kT$ is much greater than 
the energy separations between the $2s$ and $2p$ states, the reverse rate
is determined by just the ratio of statistical weights, i.e.
\begin{equation}
C_{ps} = {g_{2s}\over g_{2p}} C_{sp} = {1\over 3} C_{sp} .
\end{equation}

Similar considerations apply to the $2p$ states.  The fraction of
recombinations reaching $2p$ is $(1-f)$.  
[It is worth noting that 
processes such as Lyman-$\beta$ absorption, which ultimately produce the
$2s$ state through $1s\rightarrow 3p\rightarrow 2s$, have already been
taken into account in the calculation of $f$ \citep{spi51}.]
Radiative decay occurs through
both spontaneous Lyman-$\alpha$ emission 
(with $A_{21} = 6.25 \times 10^8$ sec$^{-1}$),
as well as stimulated emission.  We must also include radiative 
excitation from the ground state via Lyman-$\alpha$ absorption.  
Including collisions which couple to the $2s$ states, the rate
equation for $2p$ then is
\begin{eqnarray}
(1-f)\alpha n_i n_e + C_{sp}n_i n_{2s} + c\int n_\nu^{(1s)} d\nu
\int B_{12}(\nu-\nu^\prime)
n_{\nu^{\prime}} d\nu^\prime \nonumber
\\ = A_{21}n_{2p} + C_{ps}n_i n_{2p} 
+ c\int n_\nu^{(2p)}d\nu\int B_{21}(\nu-\nu^\prime) n_{\nu^{\prime}} 
d\nu^\prime ,
\end{eqnarray}
where $n_\nu^{(1s)}$ and $n_\nu^{(2p)}$ are the densities of atoms in
the $1s$ and $2p$
states per radial velocity interval, measured in frequency units; and 
$n_{\nu^\prime}$ is the photon density per frequency interval (in cm$^{-3}$ Hz$^{-1}$).
Now,
\begin{eqnarray}
B_{21}(\nu-\nu^\prime) = {c^2 \over {8\pi \nu_L^2}}A_{21} L(\nu - \nu^\prime) 
\\ {\rm and}\ \ B_{12}(\nu-\nu^\prime) = 3 B_{21}(\nu-\nu^\prime) ,
\end{eqnarray}
where $\nu_L$ is the Lyman-$\alpha$ frequency, and $L(\nu - \nu^\prime)$ 
is the Lorentz line profile.  For a thermal gas,
\begin{equation}
n_\nu^{(1s)} = {n_{1s}\over{\sqrt{\pi}\Delta\nu_D}}
e^{-{{(\nu-\nu_0)^2}\over{(\Delta\nu)_D^2}}} ,
\end{equation}
where $(\Delta\nu)_D$ is the Doppler width.  For $T = 10^4$ K,
$(\Delta\nu)_D = 1.29 \times 10^{11}$ Hz.  Clearly thermal widths will
completely dominate the natural (Lorentz) line width of Lyman-$\alpha$,
and thus we can replace $L(\nu - \nu^\prime)$ by the Dirac delta
function.  We then obtain for the absorption term:
\begin{equation}
c\int n_\nu^{(1s)} d\nu \int B_{12}(\nu-\nu^\prime)
n_{\nu^{\prime}} d\nu^\prime = {{3c^3}\over{8\pi\nu_L^2}}A_{21}
{n_{1s}\over{\sqrt{\pi}(\Delta\nu)_D}}\int n_\nu 
e^{-{{(\nu-\nu_0)^2}\over{(\Delta\nu)_D^2}}} d\nu .
\end{equation}
Of course, a similar calculation can be carried out for the stimulated
emission term.

Because the nebula is optically thick in Lyman-$\alpha$, $n_\nu$ must be 
considered carefully.  Were escape from the nebula the primary removal 
mechanism, then a steady state would result in which 
photons are created near line center, diffuse in frequency through resonant 
scattering and are
effectively removed far out in the wings.  The decrease in both creation
rate and diffusion rate with frequency offset results in a nearly flat
distribution ($n_\nu$) out to the approximate frequency at which photons 
freely escape, beyond which it drops sharply \citep{cap66}.  We write this frequency offset as 
$w(\Delta\nu)_D$, where $w$ is a dimensionless parameter expressing the offset in terms of the 
Doppler width.  We shall therefore assume
that $n_\nu = n_{Ly\alpha}/[2w(\Delta\nu)_D]$ for 
$\vert\nu - \nu_D\vert < w (\Delta\nu)_D$, and $n_\nu = 0$ for 
$\vert\nu - \nu_D\vert > w (\Delta\nu)_D$, where $n_{Ly\alpha}$ 
is the density of Lyman-$\alpha$ photons.  The $2p$ rate equation then
becomes
\begin{equation}
(1-f)\alpha n_i n_e + C_{sp}n_{2s}n_i + 
{{3 c^3}\over{16\pi\nu_L^2}} {A_{21}\over{(\Delta\nu)_D}}
{{\rm erf}(w)\over w} n_{Ly\alpha} n_{1s} = A_{21}n_{2p} .
\end{equation}
Spontaneous emission completely dominates the depopulation of the $2p$
state and thus the other two terms that appeared on the right hand side
of equation (3) have been dropped.
Stimulated emission is negligible in 
comparison with spontaneous emission for any 
reasonable value of $n_{Ly\alpha}$.  Indeed, equality of spontaneous and
stimulated emission would imply a radiation pressure (due to Lyman-$\alpha$) 
many orders of magnitude in excess of the thermal gas pressure.  Also,
the $2p \rightarrow 2s$ collision rate is negligible for any reasonable
value of $n_i$.  [We include these collisions terms insofar as they populate
the $2s$ state, however.  See equation (1).]

As we shall see, it is quite plausible that Lyman-$\alpha$ photons 
are absorbed by dust before 
significant frequency diffusion occurs.  In this case $n_\nu$ will 
simply reflect the
thermal distribution of atoms.  The resulting $2p$ rate equation is
identical to that given above provided we replace 
${{\rm erf}(w)/ w}$ with $2\sqrt{2/\pi}$.

At low densities, Lyman-$\alpha$ photons are created at the rate 
at which recombinations lead to $2p$ states, 
i.e. $(1-f)\alpha n_i n_e$.  At high densities ($n_i > 10^4$ cm$^{-3}$)
$2s$ states may be collisionally converted to $2p$ states leading to 
additional Lyman-$\alpha$ photons, and thus the Lyman-$\alpha$ creation
rate could be as large as $\alpha n_i n_e$, the total recombination rate.
We therefore define the Lyman-$\alpha$ lifetime as
\begin{equation}
t_{Ly\alpha} = {n_{Ly\alpha}\over{r\alpha n_i n_e}} ,
\end{equation}
where $1-f \le r \le 1$.  Since $2p$ states decay to the ground
state much faster than they could be collisionally converted to $2s$,
all recombinations to $2p$ are regarded as producing a Lyman-$\alpha$
photon.  Thus, $r$ could never be smaller than $1-f$.

We define
\begin{equation}
S = {{3 c^3}\over{16\pi\nu_L^2}} {A_{21}\over{(\Delta\nu)_D}}
{r\over{(1-f)}}{{\rm erf}(w)\over w}\chi n_H t_{Ly\alpha} ,
\end{equation}
where $\chi = n_{1s} / n_H$, and $n_H$ is the total hydrogen density
(atomic plus ionized).  The rate equations [(1) and (8)] can
then be solved for the $2s$ and $2p$ populations:
\begin{equation}
n_{2s} = {{\alpha n_i n_e\big[f A_{21} + (1-f)(1+S)C_{ps}n_i\big]}
\over{A_{21}(A_{2\gamma} + C_{sp}n_i)-C_{sp}C_{ps}n_i^2}}
\end{equation}
\begin{equation}
n_{2p} = {{\alpha n_i n_e\big[(1-f)(1+S)A_{2\gamma} + (1+S-
fS)C_{sp}n_i\big]}
\over{A_{21}(A_{2\gamma} + C_{sp}n_i)-C_{sp}C_{ps}n_i^2}} .
\end{equation}
The dimensionless parameter $S$ gives the rate at which $2p$ states
are produced by captured Lyman-$\alpha$ radiation in terms of the $2p$
creation rate from recombination (at low densities).

The ratio $n_{2p}/(3 n_{2s})$ provides a determination of the relative 
importance of Lyman-$\alpha$ pumping of the $2p$ states and whether a
fine structure line is expected to appear in absorption or (stimulated)
emission.  If $n_{2p}/(3 n_{2s}) > 1$, then the 9.9 GHz, 
$2s_{1/2}$-$2p_{3/2}$ line will appear in emission, and the 1.1 GHz,
$2s_{1/2}$-$2p_{1/2}$ line will appear in absorption.  Of course,
$n_{2p}/(3 n_{2s}) < 1$ implies the opposite.  (This assumes that the
two $2p$ states are populated according to their relative statistical
weights.  To evaluate cases in which $n_{2p}/(3 n_{2s})$ is of order unity, it 
would be necessary to 
separately account for the rates at which $2p_{1/2}$ and
$2p_{3/2}$ states are created and destroyed, including the collisional rates
coupling these states.)  
From equations (11)
and (12) it follows that $n_{2p}/(3 n_{2s}) > 1$ if $S > S_{crit}$
where,
\begin{equation}
S_{crit} = {{3 f}\over{(1-f)}}{{A_{21}}\over{A_{2\gamma}}} = 1.14 \times
10^8 .
\end{equation}
(The numerical value of $S_{crit}$ corresponds to $f = 1/3$.)  Because the 
$2p$ states naturally decay about $10^8$ times faster than the
$2s$ states, population equality thus
requires a pumping rate some 8 orders of
magnitude faster than the approximate rate at which $2s$ and $2p$
states are formed through recombination.

\section{The Lyman-$\alpha$ Density in HII Regions}

Determination of the Lyman-$\alpha$ density in HII regions is a
complicated transfer problem.  It is likely, however, that the dominant
mechanism for removing Lyman-$\alpha$ photons is quite straightforward,
i.e. absorption by dust \citep{kap70,spi78}.  Thus, we eschew the noncoherent
radiative transfer problem and find the upper limits to $t_{Ly\alpha}$ 
and $S$ set by absorption.  Other competing removal processes  
would reduce the lifetime, and therefore density, of 
Lyman-$\alpha$ photons, resulting in a lower value for $S$.

Using a silicate-graphite model 
for dust in HII regions \citep{aan89}, with a dust-to-gas ratio of 0.009,
the extinction at Lyman-$\alpha$ can be shown to be $N_H/(5.4 \times
10^{20} {\rm cm}^{-2})$ mag, where $N_H$ is the column density of
hydrogen.  The albedo for this mixture is about 0.4 at Lyman-$\alpha$ 
\citep{dra84}.  The lifetime of Lyman-$\alpha$ photons against absorption 
by dust can then be calculated as $t_{Ly\alpha} = (3.3 \times 10^{10}\ 
{\rm cm}^{-3}\ {\rm s}) / n_H$.  We then find that 
\begin{equation}
S = 4.2 \times 10^7\chi {r\over{(1-f)}}{{\rm erf}(w)\over w}\ .  
\end{equation}
Since $\chi << 1$ throughout
most of the volume of an HII region \citep{ost89}, we conclude that $S <<
S_{crit}$ and that the $2s$ state is overpopulated relative to the
$2p$ states.

In the harsh environments of
planetary nebulae dust might be destroyed by shocks or hard UV
radiation, or possibly separated from the ionized gas by radiation
pressure \citep{nat81,pot87}. Abundance measurements indicate, however,
that various heavy elements are depleted from the gas phase in the
ionized regions of NGC 7027 \citep{kin97} and NGC 6445 \citep{van00},
suggesting that dust has not been destroyed in significant quantities.
Additionally, planetary nebulae frequently exhibit a mid-IR spectral
component characteristic of warm dust heated by the intense radiation
field within the ionized region \citep{kwo80,hoa90,hoa92}. Summarizing
these results, \citet{bar93} has argued that dust is a common constituent
of the ionized zones of planetary nebulae, albeit with dust-to-gas 
ratios about an order of magnitude or more below that of the general
ISM.

\citet{mid90} has modeled NGC7027 and finds an extinction optical depth
of about 0.17 at 500.7 nm.  For a uniform model the column
density of hydrogen in the ionized zone is about $3.5 \times 10^{21}$
cm$^{-2}$ \citep{tho70}, along the radius of the nebula.  For
the graphite dust model used by \citet{mid90} the extinction at Lyman-$\alpha$ then
is $N_H/(1.3 \times 10^{22} {\rm cm}^{-2})$ mag; about a factor of 20
smaller than that in a general HII region described above, and roughly 
consistent with the smaller dust-to-gas ratio ($\approx 7 \times 10^{-4}$)
in this object \citep{bar93}.  For an
albedo of 0.4, the Lyman-$\alpha$ lifetime is 
$t_{Ly\alpha} = (7.6 \times 10^{11} {\rm cm}^{-3}\ {\rm s}) / n_H$, and
thus 
\begin{equation}
S \approx 9.8 \times 10^8\chi {r\over{(1-f)}}{{{\rm erf}(w)}\over w} .
\end{equation}  
If we assume that absorption by dust is the dominant process limiting the 
Lyman-$\alpha$ density, make the replacement 
${\rm erf}(w)/w \rightarrow 2\sqrt{2/\pi}$, and let $r \approx 1$, we obtain
$S = 2.3 \times 10^9 \chi$.  Since it is quite unlikely
that $\chi$ is as large as 0.05, as required for $S \approx S_{crit}$,
we conclude also in this case that the $2p$ states are underpopulated relative to the $2s$ states.

Of course, the above estimates are upper limits to the Lyman-$\alpha$
lifetime, as other processes (described in Section 1) may 
contribute to the removal of these photons.  Thus, values of $S$ estimated 
in this way are upper limits.
In addition, if escape is important then the
relevant value of ${\rm erf}(w)/w$ would be smaller than that
substituted above.

It should also be noted that Lyman-$\alpha$ radiation contributes to 
heating the dust in the ionized region.  The energy absorbed is
then reradiated in the IR.  In the case of NGC 7027 a mid-IR spectral 
component is evidently due to dust with temperature 
$T_d \approx 230$ K comixed with the 
ionized gas \citep{kwo80}.  If the heating is dominated by Lyman-$\alpha$
radiation, then
\begin{equation}
n_{Ly\alpha}h \nu_L c Q_{abs}(Ly\alpha ) = 4\left<Q(a,T_d)\right>\sigma T_d^4 ,
\end{equation}
where $Q_{abs}(Ly\alpha )$ is the absorption efficiency at 
Lyman-$\alpha$ and $\left< Q(a,T_d)\right>$ is the Planck-averaged
emissivity \citep{dra84}, a function of grain size $a$ and temperature $T$.  
Large grains, being efficient emitters in the
IR, will yield a larger estimate of $n_{Ly\alpha}$.  Thus, we assume 1
$\mu$m graphite grains for which $Q_{abs}(Ly\alpha) \approx 1$ and
$\left< Q(1\ \mu{\rm m},230\ {\rm K})\right> \approx 0.05$ \citep{dra84}.
Then,
\begin{equation}
n_{Ly\alpha} = {{4\sigma T^4}\over{h\nu_L c}}
{{\left< Q(1\ \mu{\rm m},230\ {\rm K})\right>}\over{Q_{abs}}} = 6.5
\times 10^4\ {\rm cm}^{-3} .
\end{equation}
We then find
\begin{equation}
S = {{3 c^3}\over{16\pi\nu_L^2}} {A_{21}\over{(\Delta\nu)_D}}
{1\over{\alpha(1-f)}}{{\rm erf}(w)\over w}{\chi\over{(1-\chi)}}
{{n_{Ly\alpha}}\over{n_e}} = 2.6 \times 10^{10} {{{\rm erf}(w)}\over w}
{\chi\over{(1-\chi)}} ,
\end{equation}
where we have taken $n_i = (1-\chi)n_H$ (since the fraction of atoms in 
excited, bound states is negligible).  To obtain the numerical value
given above we used the recombination coefficient in
the density bounded case with $T = 10^4$ K \citep{ost89}.
Assuming Lyman-$\alpha$ radiation is primarily 
removed through absorption by dust (in which case ${\rm erf}(w)/w$
is replaced by $2\sqrt{2/\pi}$), then  $S < S_{crit}$ unless $\chi$ exceeds
$2.4 \times 10^{-3}$, which is unlikely \citep{ost89}.  It should also be noted that
other sources, such as continuum radiation, are likely important in heating 
the dust, thereby reducing further the implied value of $n_{Ly\alpha}$ (and 
therefore $S$).  Finally, the $1\ \mu$m grain size assumed here is probably an overestimate implying that $n_{Ly\alpha}$ is also overestimated.

\section{Radiative Transfer of the Fine Structure Lines}

Evidently, the $2s$ state is overpopulated relative to $2p$, 
thus the fine structure transitions will proceed from $2s_{1/2}$ to
$2p_{3/2}$ via absorption and to $2p_{1/2}$ via stimulated emission.  
Although the $2p$ populations are probably negligible, they will be
included in the radiative transfer calculation.  The distribution of
$2p$ states between $2p_{1/2}$ and $2p_{3/2}$ may deviate somewhat from the
statistical weights ($1/3$ and $2/3$, respectively), in part because the
separate collisional rates from $2s$ are not proportional to the
statistical weights.  Since the $2p$ population is most likely
negligible, a detailed calculation of the distribution between $2p_{1/2}$ and
$2p_{3/2}$ states will not be carried out here.  Rather,
the fractional populations of $2p_{1/2}$ 
and $2p_{3/2}$ will be parameterized as $\beta_a/3$ and $2\beta_b
/ 3$, respectively.  If $\beta_a = \beta_b = 1$, then these states are
populated according to their statistical weights.  There is the obvious
constraint that $\beta_a/3 + 2\beta_b / 3 = 1$.

The fine structure transitions are allowed electric dipole transitions and the 
corresponding rates may be computed in a straightforward manner
\citep{bet57}, giving $A_a = 1.597 \times 10^{-9}$ sec$^{-1}$ 
($2s_{1/2}$--$2p_{1/2}$) and $A_b = 8.78 \times 10^{-7}$ sec$^{-1}$ 
($2p_{3/2}$--$2s_{1/2}$).  The absorption coefficient 
(valid for either transition) is
\begin{equation}
\kappa_\nu = \pm {c^2\over{8\pi \nu^2}}{g\over g_{2s}} A_\nu \big( n_{2s} -
{\beta\over 3}n_{2p}\big) ,
\end{equation}
where $g$ is the degeneracy of the final state (2 for $2p_{1/2}$, 4 for
$2p_{3/2}$; $g_{2s} = 2$).  The $-$ sign corresponds to transitions to
$2p_{1/2}$; the $+$ sign to transitions to $2p_{3/2}$, and $\beta$ is 
either $\beta_a$ or $\beta_b$, respectively.  Either final
state quickly decays to the ground state via Lyman-$\alpha$ with rate
$A_{21}$.  The natural line width of the $2s$--$2p$ transitions is
therefore dominated by the rapid decay of the $2p$ state and is $\Gamma =
A_{21}/ 2\pi = 99.8$ MHz.  For a Lorentzian profile, 
\begin{equation}
A_\nu = {{A(\Gamma / 2\pi)}\over{(\nu - \nu_f)^2 + (\Gamma / 2)^2}} ,
\end{equation}
where $A$ is either $A_a$ or $A_b$ and $\nu_f$ is the frequency of the
fine structure transition.  From equations (11) and (12) we find
\begin{equation}
n_{2s} - {\beta\over 3}n_{2p} = {{f \alpha n_i n_e}\over{(A_{2\gamma} +
C_{sp}n_i)}} \big(1 - {{\beta S}\over{S_{crit}}}\big) ,
\end{equation}
where we have assumed that the Lyman decay rate $A_{21}$ always
dominates the collision rate $C_{sp}n_i$.
The absorption coefficient at line center then is 
\begin{equation}
\kappa_\nu^{max} = \pm{c^2\over{8\pi \nu_f^2}}{g\over g_{2s}}
{{2 A}\over{\pi\Gamma}}{{f\alpha n_i n_e}\over{(A_{2\gamma}+C_{sp}n_i)}}
\big( 1 - {{\beta S}\over{S_{crit}}}\big) .
\end{equation}
This will be scaled according to the free-free absorption coefficient
\citep{alt60},
\begin{equation}
\kappa_\nu^{ff} \approx {{0.212 n_i n_e}\over{v^{2.1}T^{1.35}}} .
\end{equation}
The ratio of the line optical
depth (at line center), $\tau_\nu^{max}$, to the free-free continuum
optical depth, $\tau_\nu^{ff}$, then is
\begin{equation}
R = {{\tau_v^{max}}\over{\tau_\nu^{ff}}} = 
{K\over{(A_{2\gamma}+C_{sp}n_i)}}
\big( 1 - {{\beta S}\over{S_{crit}}}\big) ,
\end{equation}
where $K = -3.0 \times 10^{-4}$ sec$^{-1}$ for the 
$2s_{1/2}\rightarrow 2p_{1/2}$
transition and $K = 0.41$ sec$^{-1}$ for the $2s_{1/2}\rightarrow 2p_{3/2}$
transition.  We assumed $T = 10^4$ K, and that all ionizing photons are
captured by the HII region \citep{ost89}.

The equation of transfer is
\begin{equation}
{{d I_\nu}\over{d x}} = (-\kappa_\nu^{ff} - \kappa_\nu)I_\nu +
J_\nu^{ff} .
\end{equation}
Note that the line produces absorption or stimulated
emission only (through $\kappa_\nu$), but does not produce significant
spontaneous emission.  Thus, the emissivity is completely dominated by
the free-free emissivity, $J_\nu^{ff}$.  For an HII region uniform along
a ray path the solution is
\begin{equation}
T_b = T {{1 - e^{-(1+R)\tau_\nu^{ff}}}\over{1+R}} ,
\end{equation}
where the result has been expressed in terms of brightness temperature
$T_b$.  The line strength (in K) divided by the continuum brightness
temperature is
\begin{equation}
{{\Delta T_b}\over T_{b,cont}} = {1\over{1+R}}{{e^{\tau_\nu^{ff}} - 
e^{-R \tau_\nu^{ff}}}\over{e^{\tau_\nu^{ff}} - 1}} - 1 .
\end{equation}
In the optically thin limit ($\tau_\nu^{ff} << 1$) expanding (to second
order) gives
\begin{equation}
{{\Delta T_b}\over T_{b,cont}} =  -{1\over 2}R\tau_\nu^{ff} = -{1\over
2}\tau_\nu^{max} .
\end{equation}
This result is in agreement with that obtained by \citet{ers87} for the 
optically thin limit.
In the optically thick limit ($\tau_\nu^{ff} >> 1$),
\begin{equation}
{{\Delta T_b}\over T_{b,cont}} = {{-R}\over{1+R}} .
\end{equation}
The lines do not disappear in the optically thick limit as long as
the $2s$ and $2p$ states are not in local thermodynamic equilibrium.  (See also
Ershov 1987.)  The
$10^4$ K microwave radiation field is not sufficient to overcome the
rates described in Section 2 and thus it can not establish equilibrium.
For example, in the likely case where $n_{2p}/(3n_{2s}) << 1$, the $2p$ 
levels are drained by Lyman-$\alpha$ emission faster than the microwave
radiation field can return these states to $2s$ via either absorption or
stimulated emission.  (See also Section 2.)

Each fine structure level is split into two hyperfine levels as shown in
Figure 1.  The allowed transitions are also indicated.  Both fine structure
lines are split into three hyperfine components.  The relative
intensities of the components can be calculated from the appropriate sum
rules \citep{sob92}.  For the 910, 1088, and 1147 MHz lines the ratios
are 1:2:1.  The 9852, 9876, and 10030 MHz lines appear in the ratios
1:5:2.

\section{Prospects for Detection}

The 9.9 GHz transitions are intrinsically about three orders of magnitude 
stronger than the 1.1 GHz transitions.  The solution to the radiative transfer equation indicates that, in general, the line brightness temperature $\Delta T_b$ grows as the square of the free-free optical depth in the optically thin limit (equation 28).  One factor comes from the growth of free-free emission, which effectively forms the ``background", and the other from the
proportionate growth of the line absorption.  In the optically thick limit, the growth in line temperature saturates (equation 29).  Most HII regions of interest are optically thin at 9.9 GHz and optically thick at 1.1 GHz.  In general, the higher optical depth at 1.1 GHz does not significantly offset the relative weakness of these transitions.  We find, therefore, that the 9.9 GHz lines offer the better prospects for detection.  

Table 1 gives estimates of the line strengths for various HII
regions and components, and planetary nebulae.  It was
assumed that $S=0$, in which case the 9.9 GHz 
lines would appear in absorption and the 1.1 GHz lines in stimulated emission.  
Most of the entries in Table 1
correspond to high emission measure components in HII regions, which are
listed according to the nomenclature of the original references.  The
published emission measure values $E$ were used to calculate the 
free-free optical depth and the
continuum brightness temperature (assuming in most cases 
$T_e \approx 10^4$ K).  The line-to-continuum optical depth ratios $R$ are calculated from the published electron densities (equation 24).  Notably, our predictions for both the 1.1 GHz and 9.9 GHz lines from Orion A (M42) are in good agreement with those of \citet{ers87}.

The estimates in Table 1 take into account the distribution of the line strength over three hyperfine components making up each fine structure line and the consequent line blending.  At 9.9 GHz, the strongest line (9876 MHz) will be blended with the weakest line (9853 MHz).  The peak temperature occurs at 9874 MHz and is 75\% of that calculated using equation 26 for a single (fictitious) fine structure line.  The 10030 MHz line will be somewhat distinct with a peak temperature equal to 32\% of that predicted from equation 26, including contributions from the wings of the 9852 and 9876 MHz lines.  The situation at 1.1 GHz is similar.  The 1088 MHz line will appear blended with the 1147 MHz line, with a peak value of 
63\% at 1093 MHz.  The 910 MHz line will remain distinct with a peak value of 30\%, including line wing contributions from the other two lines.  The estimates in Table 1 give the line temperature at the peak of the brightest spectral feature in either the 9.9 GHz or 1.1 GHz blended multiplet.  At both frequencies, the resulting profile of a strong blended feature and a distinct weaker line, combined with Lorentzian line shapes, will provide a unique detection signature.  Figure 2 shows the predicted appearance of the blended multiplet around 9.9 GHz.  

\begin{figure}
\epsscale{0.85}
\plotone{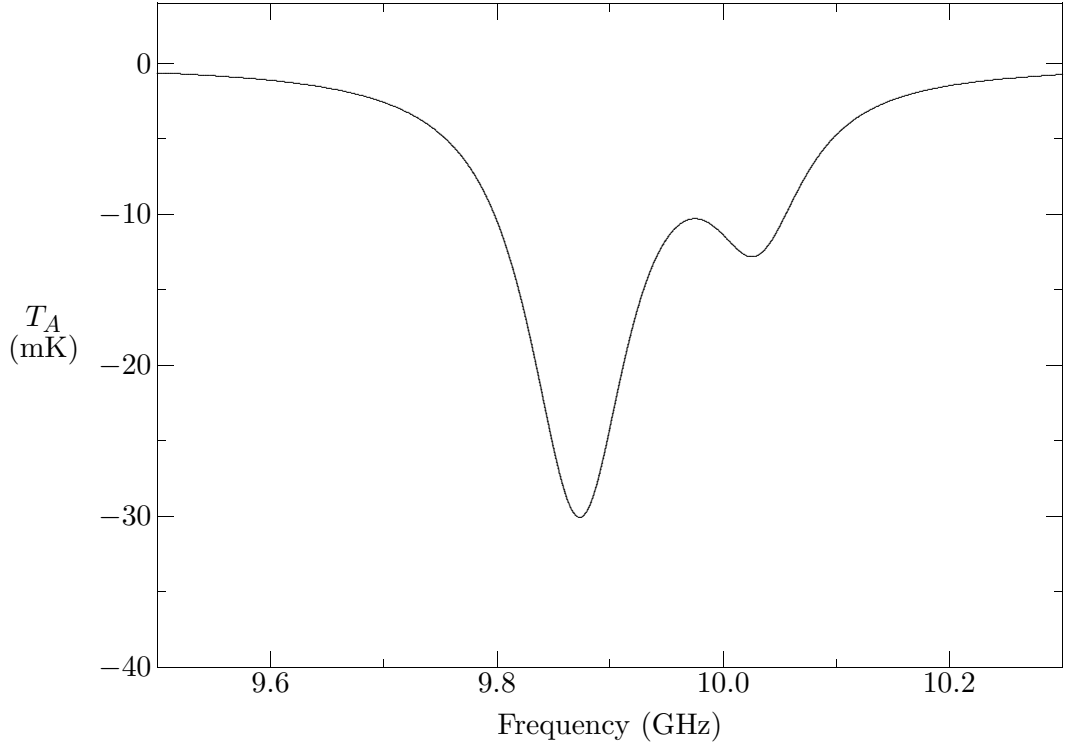}
\caption{Hyperfine multiplet near 9.9 GHz.  Three hyperfine components at 9.852, 9.876, and 10.030 GHz appear blended due to the large natural line width, 99.8 MHz, of these transitions.  The individual line strengths are in the ratio 1:5:2.  The vertical axis is the predicted antenna temperature for Green Bank Telescope observations of NGC 7027, corresponding to the shell model, listed in Table 1 as \{NGC 7027\}$_S$.  The relative populations of the $2p$ states are assumed negligible (see text), and thus the 9.9 GHz lines appear in absorption.  The line profiles are completely dominated by natural line width and thus strong Lorentzian wings are apparent.} 
\end{figure}

The Green Bank Telescope (GBT) provides a 
realistic hope of detection of the 9.9 GHz lines.  Having a clear aperture it should be
relatively free of standing waves in the antenna structure.  Thus, it should be
possible to search for very broad spectral features over bandpasses as
large as 800 MHz.  Nevertheless, the high emission measure HII regions that are likely to show the lines are typically quite
compact.  Therefore, the estimated line antenna temperatures $(\Delta T_a)_{9.9}$ include the effects of dilution in the $1^\prime.21$ GBT beam.  The estimated values of $\Delta T_a$ at 1.1 GHz were calculated using either the $11^\prime$ GBT beam (subscript $G$), or where appropriate, the Arecibo $4^\prime.3$ beam (subscript $A$).

Quite possibly the observational sensitivity will be limited by
systematic effects in which the continuum antenna temperature is
modulated by frequency dependent gain variations which are not entirely
removed by calibration procedures.  Thus the limiting factor may be the line-to-continuum ratio,
$\Delta T / T$, which is also tabulated for both lines.

The estimates given in Table 1 are only approximate.  The underlying 
observations are biased in favor of component sizes to
which various interferometer arrays are sensitive.  
In complex HII
regions structures are present on a range of scales, down to very high
emission measure arc-second scale components \citep{tur84}.  The 
components included in the table 
(typically a few tens of arc seconds in size) were selected because they 
contribute significantly to the total flux in the GBT beam.
The more compact, arc-second scale 
components typically yield larger values of $\Delta T / T$ (despite 
densities $> 10^4$ cm$^{-3}$ and consequent collisional de-excitation),
yet they tend to contribute relatively little to the total
flux in the GBT beam.  Conversely, extended, low emission measure components 
in the beam will
add to the antenna temperature while contributing little line
absorption.  It should also be noted that 
the presence of structures having a wide range of
densities implies that the uniform model considered in Section 4 is an
oversimplification.  Inhomogeneous structure, including clumping, in the emission regions
would imply that the emission measure estimates in Table 1 represent an average over the surface of the source.  In the optically thin case, i.e. most sources at 9.9 GHz, a redistribution of emission measure, and therefore optical depth, will tend to strengthen the overall estimated line strength $\Delta T_b$ due to its $\tau^2$-dependence (equation 28).  Collisional de-excitation in denser regions of the source, however, will tend to reduce the line optical depth (equation 24).  Because $R$ is density dependent, particularly for densities above about $1.5 \times 10^4$ cm$^{-3}$, the brightness temperature for a homogeneous source (equation 26) cannot be rescaled using some weighted optical depth; rather, detailed modeling would be required.  

Three high emission measure planetary nebulae are also included in the 
table (IC 418, NGC 7027, and NGC 6572).  These objects tend to have somewhat simpler structure than the
more complex HII regions.  Nevertheless, NGC 7027 has a well documented shell structure.  In this case, a number of emission line diagnostics indicate 
$n_e \approx 5 \times 10^4$ cm$^{-3}$ \citep{mid90}.  Subarcsecond radio images 
show that the emission originates
from a shell with a peak emission measure of about $2.7 \times
10^8$ pc cm$^{-6}$ \citep{bry97}.  Both results are consistent with an area 
filling factor of about 30\%, with characteristic emission measure of
about $1.2 \times 10^8$ pc cm$^{-6}$.  The higher
emission measure boosts the line-to-continuum ratio, whereas the higher
density acts oppositely due to collisional de-excitation.  The net result
in this case is a somewhat stronger estimated line with $\Delta T_b / T_{b,cont}
\approx -1.4 \times 10^{-3}$.  This case is worked out in Table 1 
as \{NGC 7027\}$_S$, and depicted in Figure 2. 

Generally, compact, high emission measure objects tend to give stronger 
line-to-continuum ratios, despite their higher densities (which result in collisional de-excitation).  This is because HII
regions tend to be loosely organized along domains of constant excitation
parameter $U$ \citep{hab79}.  Using size $d$ as a free parameter, then 
$E \propto U^3/d^2$.  For $E < 3.8\times 10^8$ pc cm$^{-6}$, the HII region is optically thin at 9.9 GHz and, for $S=0$
\begin{equation}
{{\Delta T_b}\over{T_{b,cont}}} \approx -10^{-2}{E_8\over{n_4 + 1.5}}
\end{equation}
where $E_8 = E/(10^8\ {\rm pc\ cm}^{-6})$ and $n_4 = n/(10^4\ 
{\rm cm}^{-3})$.
Thus, in the low density limit ($n_4 << 1.5$) the line-to-continuum ratio grows in direct proportion to $E$, and therefore to $d^{-2}$ for fixed $U$.  Of course, the most compact objects have densities in excess of $1.5\times 10^4\ {\rm cm}^{-3}$, in which case the 
line-to-continuum ratio increases with $d^{-1/2} \propto E^{1/4}$ for fixed $U$, since $n_e \propto (U/d)^{3/2}$.  For emission measures above $\approx 4\times 10^8$ pc cm$^{-6}$ an HII region becomes optically thick at 9.9 GHz and the advantage of increased emission measure is lost.  In such cases,
higher densities will reduce the line-to-continuum ratio.  

As discussed above, the 1.1 GHz lines are considerably weaker.  Under the 
most favorable conditions of high free-free optical depth 
and low density, we find
\begin{equation}
{{\Delta T_b}\over{T_{b,cont}}} \approx -R \approx 3.6 \times 10^{-5}\ ,
\end{equation}
assuming, as discussed above, that $S = 0$.  These conditions 
are not uncommon and exceptionally high emission measures are not required to achieve high optical depth at 1.1 GHz.
In the case of M 42 the
optical depth at 1.1 GHz is 1.2 and the beam diluted line strength would be
about 4 mK, versus a continuum antenna temperature of $\approx$ 400 K.

\section{Conclusions}

The metastable $2s_{1/2}$ state of hydrogen is likely overpopulated in HII regions.  Lyman-$\alpha$ pumping of the $2p$ states is expected to be negligible due to absorption of Lyman-$\alpha$ radiation by dust.  Thus, the $2s_{1/2}\rightarrow 2p_{3/2}$ transitions (9.9 GHz) are predicted to appear in absorption and the $2s_{1/2}\rightarrow 2p_{1/2}$ transitions (1.1 GHz) in stimulated emission.  Because of the short lifetime of the final $2p$ states, the width of the lines is dominated completely by intrinsic line width.  In effect, then, the power is distributed over $\approx 100$ MHz of line width resulting in very weak lines.  In addition, the power is distributed over three strongly blended hyperfine lines in each multiplet.

Searching for the 9.9 GHz lines in high emission measure HII regions offers the best prospects for detection.  In the optically thin limit, the line strength varies as the square of the free-free optical depth.  Predicted line-to-continuum ratios (in absorption) range up to several tenths of a percent in W58A, including the effects of line blending.  With the Green Bank Telescope, the predicted peak absorption line strength may reach $\Delta T_a\approx -170$ mK in this case, allowing for the redistribution of line strength over the three hyperfine lines.  Other high emission HII regions are expected to show somewhat weaker 9.9 GHz lines, for example, with line-to-continuum ratios of about 0.1 percent and line strengths of tens of mK with the Green Bank Telescope.  These predictions are uncertain, however, owing to biases inherent in estimating emission measures as well as selection effects in various interferometric surveys of compact HII regions.  

These conclusions apply to thermal sources.  An important extension of this work would consider the broad line regions of active galactic nuclei and quasars in which a strong nonthermal microwave radiation field could influence the populations of the $2s$ and $2p$ levels, as well as provide a background for line absorption or stimulated emission.

In general, detection of the fine structure lines of hydrogen will be challenging due to the extraordinary line width and blended structure.  The observations will require meticulous baseline calibration and subtraction.

We thank Drs. R. Brown and J. Simonetti for useful discussions, and Dr. A. Ershov for bringing his work to our attention.  Portions of this work were completed while one of the authors (B.D.) was a Visiting Scientist at the National Radio Astronomy Observatory (NRAO) in Green Bank, WV, and also a faculty member in the Department of Physics at Virginia Tech.  This work was supported by the Glaxo-Wellcome Endowment at the University of North Carolina-Asheville and by National Science Foundation grant AST-0098487 to Virginia Tech.  The National Radio Astronomy Observatory is a facility of the National Science Foundation operated under cooperative agreement by Associated Universities, Inc.


\clearpage

\begin{deluxetable}{crrrrrrrrrrrr}
\rotate
\tabletypesize{\scriptsize}
\tablewidth{0pt}
\tablecaption{HII Regions --- Estimated line strengths. \label{tbl-1}}
\tablehead{
\colhead{} & \colhead{$n_e$}   & \colhead{$E$}   &
\colhead{} &
\colhead{}  & \colhead{} & \colhead{} &\colhead{$\big(\Delta T_a\big)_{9.9}$}     
& \colhead{} & \colhead{} &\colhead{$\big(\Delta T_a\big)_{1.1}$}  & \colhead{}\\
\colhead{HII Region} & \colhead{(cm$^{-3}$)}   & \colhead{(pc cm$^{-6}$)}   &
\colhead{$\theta$} &
\colhead{$\tau_{9.9}^{ff}$}  & \colhead{$R_{9.9}$} & \colhead{$\big({{\Delta T_b}\over{T_{b,cont}}}\big)_{9.9}$} 
&\colhead{(mK)}     
& \colhead{$R_{1.1}$} & \colhead{$\big({{\Delta T_b}\over{T_{b,cont}}}\big)_{1.1}$} &\colhead{(mK)}  & 
\colhead{Reference}\\
}
\startdata
W3.1 &8200 &$1.6\times 10^7$ &$12\arcsec\times 10\arcsec$ &0.043 &0.033 &$-5.2\times 10^{-4}$ & -5.0 
&$-2.4\times 10^{-5}$ &$1.4\times 10^{-5}$ &0.04$_G$ & 1\\
W3.4 &5700 &$1.2\times 10^7$ &$20\arcsec\times 15\arcsec$ &0.032 &0.036 &$-4.2\times 10^{-4}$ & -7.5 
&$-2.7\times 10^{-5}$ &$1.4\times 10^{-5}$ &0.10$_G$ & 1\\ 
W3.5 &6100 &$1.4\times 10^7$ &$20\arcsec\times 15\arcsec$ &0.037 &0.035 &$-5.0\times 10^{-4}$ & -10.
&$-2.6\times 10^{-5}$ &$1.5\times 10^{-5}$ &0.10$_G$ & 1\\

IC 418 &$2.6\times 10^4$ &$7.7\times 10^6$ &$11.\arcsec 5$ &0.020 &0.019
&$-1.4\times 10^{-4}$ &-0.7 &$-1.4\times 10^{-5}$ &$6.0\times 10^{-6}$ &0.02$_G$ & 2\\

M42  &2200 &$3.2\times 10^6$ &$2.\arcmin 6\times3.\arcmin 5$ &0.012 &0.044& $-1.9\times 10^{-4}$ & -17. &$-3.2\times 10^{-5}$ &$9.4\times 10^{-6}$ &3.7$_G$ & 3\\

NGC6572 &$2.5\times 10^4$ &$3.2\times 10^7$ &6.\arcsec 4 &0.085 & 0.019 
&$-6.1\times 10^{-4}$ &-4.0 &$-1.4\times 10^{-5}$ &$8.8\times 10^{-6}$ &0.053$_G$ & 2\\

M17 &950 &$2.4\times 10^6$ &$4.\arcmin 5\times 1.\arcmin 8$ &0.064 &0.047 &$-1.1\times 10^{-4}$ & -7.5 &$-3.4\times 10^{-5}$ &$6.3\times 10^{-6}$ 
&2.1$_G$ & 3\\
W48 &8075 &$1.8\times 10^7$ &$11\arcsec \times 28\arcsec$ &0.048 &0.033 &$-5.8\times 10^{-4}$ &-16.
&$-2.4\times 10^{-5}$ &$1.5\times 10^{-5}$ &0.63$_A$ & 4\\

W49A.1 &5000 &$1.6\times 10^7$ &$7.\arcsec$ &0.043 &0.038 
&$-5.8\times 10^{-4}$ &-2.5 &$-2.8\times 10^{-5}$ &$1.6\times 10^{-5}$ &0.13$_A$ & 5\\
W49A.4 &7000 &$3.1\times 10^7$ &$7.\arcsec$ &0.083 &0.034 
&$-1.0\times 10^{-3}$ &-8.2 &$-2.5\times 10^{-5}$ &$1.6\times 10^{-5}$ &0.13$_A$ & 5\\
W49A.5 &4000 &$1.9\times 10^7$ &$15\arcsec$ &0.051 &0.040 
&$-7.4\times 10^{-4}$ &-16. &$-2.9\times 10^{-5}$ &$1.8\times 10^{-5}$ &0.59$_A$ & 5\\
W49A.6 &4000 &$1.4\times 10^7$ &$11\arcsec$ &0.037 &0.040 
&$-5.5\times 10^{-4}$ &-4.6 &$-2.9\times 10^{-5}$ &$1.6\times 10^{-5}$ &0.28$_A$ & 5\\

W51A.b &4800 &$1.1\times 10^7$ &$12\arcsec$ &0.029 &0.038 
&$-4.1\times 10^{-4}$ &-3.3 &$-2.8\times 10^{-5}$ &$1.4\times 10^{-5}$ &0.30$_A$ & 5\\
W51A.d &$1.2\times 10^4$ &$4.4\times 10^7$ &$7\arcsec$ &0.012 &0.028
&$-1.2\times 10^{-3}$ &-13. &$-2.1\times 10^{-5}$ &$1.3\times 10^{-5}$ &0.10$_A$ & 5\\
W51A.e &5000 &$2.1\times 10^7$ &$21\arcsec$ &0.056 &0.038
&$-7.5\times 10^{-4}$ &-37. &$-2.8\times 10^{-5}$ &$1.7\times 10^{-5}$ &1.1$_A$ & 5\\

W58A &$3.9\times 10^4$ &$3.5\times 10^8$ &$6\arcsec$ &0.931 &0.014 
&$-4.2\times 10^{-3}$ & -170. &$-1.0\times 10^{-5}$ &$6.3\times 10^{-6}$ &0.04$_A$ & 6\\

DR 21&$1.2\times 10^4$ &$3.4\times 10^7$ &$20\arcsec\times 7\arcsec$ &0.091 &0.028 &$-1.0\times 10^{-3}$ &-23.
&$-2.1\times 10^{-5}$ &$1.3\times 10^{-5}$ &0.04$_G$ & 6\\

NGC 7027 &$1.9\times 10^4$ &$4.2\times 10^7$ &$11.\arcsec 5$ &0.11 &0.023 
&$-1.0\times 10^{-3}$ &-25. &$-1.7\times 10^{-5}$ &$1.0\times 10^{-5}$ &0.03$_G$ & 2\\
{{\{NGC 7027\}}$_S$}\tablenotemark{a} &$5.0\times 10^4$ &$1.3\times 10^8$ &$17.\arcsec 3$ &0.34 &0.012
&$-1.4\times 10^{-3}$ &-30. &$-8.6\times 10^{-5}$ &$5.4\times 10^{-5}$ &0.05$_G$ & 7\\
\enddata
\tablenotetext{a}{Shell model for NGC 7027.  See text.}
\tablecomments{\ Antenna temperatures ($\Delta T_a$) include the effects of beam dilution with the Green Bank Telescope at 9.9 GHz and 1.1 GHz (indicated by subscript $G$), and Arecibo at 1.1 GHz (subscript $A$).}
\tablerefs{(1) Webster \& Altenhoff 1970; (2) Thomasson \& Davies 1970; (3) Schraml \& Mezger 1969; (4) Krassner et al. (1983); (5) van Gorkom et al. (1980); (6) Wynn-Williams (1971); (7) Bryce et al. 1997
}
\end{deluxetable}
\clearpage
\end{document}